\documentclass{aa}
\usepackage{epsfig,amsmath,amssymb}

\def\msol{M_\odot}

\def\mv{M_{\rm V}}

\def\te{T_{\rm eff}}

\def\simgr{\,\hbox{\hbox{$ > $}\kern -0.8em \lower 1.0ex\hbox{$\sim$}}\,}
\def\simle{\,\hbox{\hbox{$ < $}\kern -0.8em \lower 1.0ex\hbox{$\sim$}}\,}
\def\beq{\begin{equation}}
\def\eeq{\end{equation}}

\begin{document}

\title{Nonradial oscillations in classical Cepheids: the problem revisited}

 \author{C. Mulet-Marquis\inst{1}, W. Glatzel\inst{2}, I. Baraffe\inst{1,2}, C. Winisdoerffer\inst{1}
}

\offprints{C. Mulet-Marquis}

\institute{C.R.A.L, CNRS, UMR5574
 \'{E}cole normale sup\'erieure, 46 all\'ee d'Italie, 69007 Lyon, France (cedric.mulet-marquis, ibaraffe, cwinisdo@ens-lyon.fr)
\and
Institut f\"ur Astrophysik, Georg-August-Universit\"at G\"ottingen, Friedrich-Hund-Platz 1, D-37077 G\"ottingen (wglatze@astro.physik.uni-goettingen.de)
}

\date{Received /Accepted}

\titlerunning{Nonradial oscillations in classical Cepheids}
\authorrunning{Mulet-Marquis et al.}
\abstract{We analyse the presence of nonradial oscillations in Cepheids, a problem which has not been theoretically revised since the work of Dziembowsky (1977) and Osaki (1977). Our analysis is motivated by a work of Moskalik et al. (2004) which reports the detection of low amplitude periodicities in
 a few Cepheids of the large Magellanic cloud. These newly discovered periodicities
were interpreted as nonradial modes.}
{Based on linear nonadiabatic stability analysis, our goal is to reanalyse the presence and stability of nonradial modes, taking into account improvement in the main input physics required for the modelling of Cepheids.}
{We compare the results obtained from two different numerical methods used to solve
the set of differential equations: a matrix method and the Ricatti method.}
{We show the limitation of the matrix method to find low order p-modes ($l<6$), because of their dual character  in evolved stars such as Cepheids. For higher order
p-modes, we find an excellent agreement between the two methods. }
{No nonradial instability is found  below $l=5$, whereas many unstable nonradial modes
exist for higher orders. We also find that nonradial modes remain unstable,
even at hotter effective temperatures than the blue edge of the Cepheid 
instability strip, where no radial pulsations are expected.}


\maketitle

\section{Introduction}

Cepheids are well known radial pulsators, oscillating in the  fundamental mode or the low overtones. Despite a very abundant literature analysing theoretically and observationally their radial pulsation properties, only very few studies have been devoted to nonradial oscillations in these stars. 
Dziembowski (1971) was the first to study nonradial oscillations in Cepheids, based on a quasiadiabatic linear stability analysis, and found that low-order modes ($l \le 2$) are stable.  Following this study, Osaki (1977) and Dziembowski (1977) showed that higher degree nonradial modes ($l \ge 4$) can become unstable. These analysis highlighted the difficulty to calculate low-order p-modes in such evolved stars. These modes show a dual character as they can also propagate in the gravity-wave region, behaving as high-order gravity modes in the central regions and thus having several thousand nodes there.
Since no observational evidence for the presence of nonradial modes in Cepheids was available at the time of these studies, no further theoretical analysis of the stability of nonradial modes was performed. 

Later on, however, due to advances in spectroscopy and radial velocity measurement 
 techniques, the presence of nonradial modes in classical Cepheids was more convincingly  considered to explain anomalous behaviors such as amplitude modulations (Hatzes \& Cochran 1995; 
Van Hoolst \& Waelkens 1995; Koen 2001; Kovtyukh et al. 2003). 
These have been interpreted either as the beating of two linearly unstable radial and nonradial modes with similar frequencies, or as a nonlinear interaction between a linearly unstable radial mode and a linearly stable, low degree, low-order nonradial mode (Van Hoolst \& Waelkens 1995; Van Hoolst et al. 1998). The latter hypothesis was theoretically investigated by
Van Hoolst \& Waelkens (1995), based on frequencies and linear growth rates
of the nonradial modes calculated previously by Osaki (1977). 

More recently, 
Moskalik et al. (2004) found low amplitude secondary periodicities in a few first overtone Cepheids of the Large Magellanic Cloud, with frequencies close to the unstable, radial, first overtone. These newly discovered periodicities were interpreted as nonradial modes. They provide, to our knowledge, the first direct evidence for the presence of nonradial modes in classical Cepheids. But unfortunately, these observations have not been confirmed yet. The observations of Moskalik et al. (2004) are based on photometrical data, implying that if  nonradial modes are indeed detected,
they must be low-degree modes. According to Osaki (1977),
however, such modes should be stable. Motivated by the observations of Moskalik et al. (2004) and the fact that since the work of Osaki (1977), no updated stability analysis of nonradial modes has been performed, we reinvestigate this problem, taking into account the improvement in the main input physics (opacities, equation of state) required for the modelling of Cepheids.

We have used two different numerical methods for solving the set of linear pulsation equations. The first one is based on a Henyey-type relaxation scheme, which is the most commonly used in the community; the second one is based on the Riccati method (see Gautschy \& Glatzel, 1990, and references therein). The two methods are briefly described in \S 2 and \S3 respectively. 
Because of the above-mentioned dual character of low-degree p-modes in highly evolved stars, the first method requires an extremely large number of grid-points  
for the Cepheid model because of the large number of nodes in the central region. 
Such method thus suffers from resolution problems and a lack of accuracy. One of our purposes is to analyse and highlight such limitations and is done in \S2. To 
overcome the problems, several ideas have been suggested, such as asymptotic methods providing analytical solutions  (Dziembowski 1977;
Lee 1985). Osaki (1977) found that the stellar envelope could be regarded as a unique pulsating unit, with an appropriate boundary condition at the bottom of the envelope which in principle, should take into account the enery leakage into the core. 
As a more attractive alternative, the Riccati method as used by Gautschy \& Glatzel (1990), avoids the extra-complexity of asymptotic methods and can  
take into account the entire stellar structure, providing a correct  description of the energy leakage into the core without resorting to any approximation. 

In section \S 4 we present the outcome of our stability analysis and compare the results obtained with both numerical methods for low-degree p-modes. 

\section{Linear stability analysis based on the Henyey-method}

\subsection{Linear pulsation equations}
The pulsation calculations are performed with a nonradial code originally developed by Lee (1985) and previously used by one of the authors for an extensive study of radial pulsations in classical Cepheids (Alibert et al. 1999).   We briefly describe the method below. 
The system of linearised equations describing small amplitude stellar pulsations can be found
   in Unno et al. (1989). The eigenfunctions $X(r,\theta,\phi,t)$, solutions of this system of equations, are expressed in terms of spherical harmonics as 
   $X = x(r) Y^l_m(\theta,\phi)e^{i \sigma t}$, where $l$ indicates the degree of the eigenmode and $m$ its azimuthal number. $\sigma = \sigma_r + i \sigma_i$ is the eigenfrequency of the mode, with $P=2 \pi /\sigma_r$ the pulsation period and
 $\sigma_i$ characterising the stability of the eigenmode. Positive values of $\sigma_i$ indicate stable modes. In the following, all frequencies are given in units of
$\sigma_0= \sqrt{\dfrac{Gm}{R^3}}$ , with $R$ the radius and $m$ the mass of the star.
Convection is frozen in, assuming that the perturbation of the convective flux  in the linearised energy conservation equation is neglected. Though crude, this approximation can be justified as  we will be mainly interested in Cepheids close to the blue edge of the instability strip, where convection plays a minor role. As shown in Alibert et al. (1999), frozen-in convection
provides a theoretical blue edge in good agreement with observed radial fundamental  and first overtone Cepheid pulsators.
The boundary conditions are those imposed by the regularity of the eigenfunctions (see Unno et al. 1989). A normalisation condition is added : the radial component of the displacement is set to one  at the surface of the star.
We adopt, as in Unno et al. (1989), the following variables $y_1 = \dfrac{\xi_r}{r}$,
$y_2 = \dfrac{1}{gr}\left( \dfrac{p'}{\rho}+\phi '\right)  = \dfrac{\sigma^2 r}{g}\dfrac{\xi_h}{r}$,
$y_3 = \dfrac{1}{gr} \phi '$, $y_4 = \dfrac{1}{g}\dfrac{\mathrm{d} \phi '}{\mathrm{d} r}$,
$y_5 = \dfrac{\delta S}{ C_p}$ and
$y_6 = \dfrac{\delta L_{\mathrm{rad}}}{L_{\mathrm{rad}}}$,
   with $ \xi_r$ the radial and $ \xi_h$ the horizontal component of the
   displacement respectively,
   $g$ the gravitational acceleration, $ \phi$ the gravitational potential, $p'$ and $\phi'$ 
  the Eulerian perturbations of the pressure and gravitational potential, $ L_{\mathrm{rad}}$ and  $\delta L_{\mathrm{rad}}$
   the radiative luminosity and its Lagrangian perturbation,
   $  \delta S$  the Lagrangian perturbation of the entropy,
  and  $ C_p$ the specific heat at constant pressure.
  The inner  boundary conditions are:
\beq
\frac{c_1 \sigma ^2}{l}y_1 - y_2 =0  \\
\eeq
and 
\beq
ly_3-y_4=0, 
\eeq
with $ c_1 = \frac{m}{m_{\rm r}}\frac{r^3}{R^3}$ and $m_{\rm r}$ is the mass inside the sphere of radius $r$. The outer boundary conditions at $r$=R are:
\beq
(l+1)y_3+y_4=0, 
\eeq
\beq
(2-4\nabla_{  \mathrm{ad}}V)y_1+4\nabla_{\mathrm{ad}} V(y_2-y_3) + 4y_5-y_6=0,
\eeq
\beq
y_1=1.
\eeq
with $ \nabla_{\mathrm{ad}}$ the adiabatic gradient and
 $V = -\dfrac{\mathrm{d} \, \ln P}{\mathrm{d} \, \ln \, r}$.
Finally, the system of linearised equations is numerically solved with a Henyey-type relaxation method. It requires a guess for the eigenfrequency which is derived from the solution of the adiabatic problem.

\subsection{  Modal classification }

\begin{figure}
\psfig{file=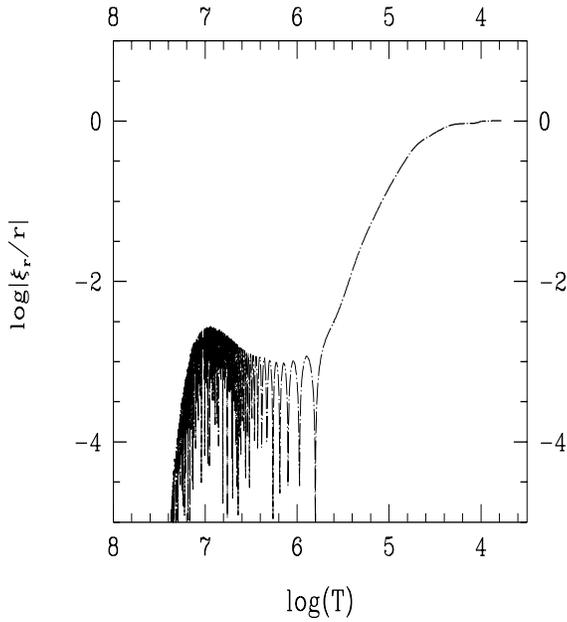,height=120mm,width=88mm} 
\caption{Radial displacement as a function of temperature of a p-mode with $l$=10, frequency $\sigma_r=3.62$ in units of $\sigma_0$, for a typical Cepheid model ($m=5M_\odot$, $\log L/L_\odot=3.1$, $T_{\mathrm{eff}}=5930 \, \mathrm{K}$).}
\label{fig1}
\end{figure}

The standard modal classification of nonradial modes, based
on  the determination of the number of nodes $N_{\rm g}$ in the gravity-wave and $N_{\rm p}$ in the acoustic-wave zones respectively (see e.g Unno et al. 1989) usually fails for evolved stars, as underlined by Dziembowski (1971), because of the dual character of the modes. Modes with p-mode character in the envelope do rapidly oscillate,
like g-modes, in the central region. 
This behavior is illustrated in Fig. \ref{fig1} 
which displays  the radial displacement of a p-mode of degree $l$=10 found in a typical Cepheid model. In this case  the number of nodes calculated with the present numerical method is inaccurate and  meaningless, with  $N_{\rm p}$=117 and $N_{\rm g}$=143. Another method must be adopted to select and classify p-modes, which are the most interesting ones since they can propagate to the surface and are thus potentially detectable.

Also, a huge number of solutions is found, among which some are unphysical and result from numerical
noise. For the selection of physical p-modes, we have used a criterion based on the modal kinetic energy, $E_{\rm c}$, which reads  for a layer between radii $r$ and $r+\mathrm{d}r$:

\begin{displaymath}
 E_{\rm c} = \frac{1}{2} \sigma^2  \rho r^2 |\delta r | ^2,
\end{displaymath} 
with
$ |\delta r | ^2 = |\xi_r|^2 + l(l+1)|\xi_h|^2$.

\begin{figure}
\psfig{file=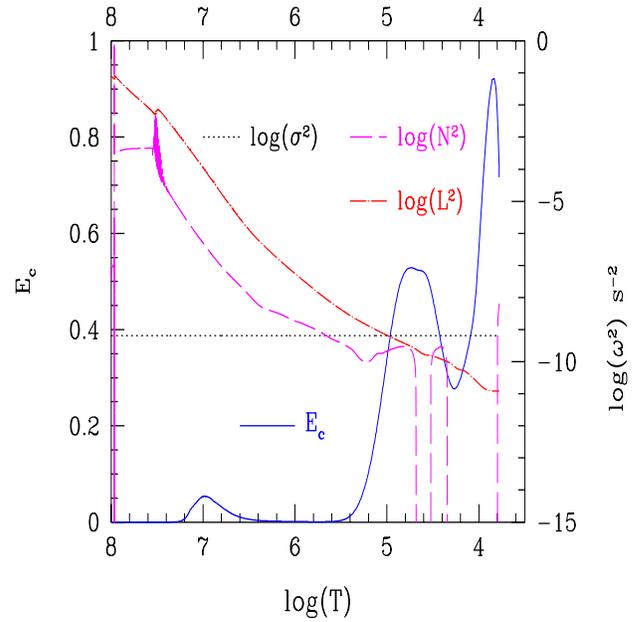,height=120mm,width=88mm}
\caption{ Kinetic energy $E_{\rm c}$ (solid line, arbitrary units) for the same eigenmode with $l$=10 
as in Fig. \ref{fig1}.
$E_{\rm c}$ reaches its maximum value  in the p-zone, where the eigenfrequency (dotted line) is greater than the
 Brunt-V\"{a}is\"{a}l\"{a} frequency $N$ (dashed line)
and greater than the Lamb frequency $L$ (dash-dotted line). All frequencies are
in s$^{-1}$  (right-hand $y$-axis). 
This eigenmode fullfills the selection criterion of  a p-mode (see text \S 2.2).}
\label{fig2} 
\end{figure}

If the kinetic energy of an eigenmode reaches its maximum value in the acoustic-wave zone
($\sigma > N, L$, with $N$ and $L$ being the Brunt-V\"{a}is\"{a}l\"{a} and the Lamb frequencies respectively),
 it is  selected as a correct solution, which is illustrated in Fig. \ref{fig2}.  Each selected eigenmode, for a given $l$, is then classified
 as fundamental  mode F or overtone (1H, 2H, {\it etc}) with increasing value of $\sigma_r$.
 This selection method allows us to find  p-modes with high-degree, $l \ge 6$, but fails for lower
 degrees due to the limitation of the numerical method. Although we  find numerous solutions for $l < 6$,  many of them have very close  values of $\sigma_r$ but very different values of $\sigma_i$.
Moreover, they all have pathological eigenfunctions and 
a selection of  the correct solution based on the shape of the eigenfunctions or the kinetic energy is not possible. 
  
Since the problems above mentioned stem from the rapid oscillations of  the eigenfunctions close to the
center of the star, one could think of using the quite common ``envelope model" approach,
namely solving the dispersion equation using envelope models rather than complete stellar models.
This approach, however,  is not useful in the present context since one needs to
 transfer the unambiguous inner boundary conditions taken at $r=0$ 
to the inner edge of the envelope. 
Since  there is no zone where the evanescent oscillation would be rapidly switched off, it is quite difficult to choose the correct inner boundary conditions 
with envelope models.
In order to get rid of this problem, the inner boundary conditions should be taken deep enough, if not, they will affect the solution.  
We found that using envelope models requires the same typical resolution as the one required with complete stellar models. There 
is thus no advantage of using an envelope model rather than a complete one since the size of the matrix which needs to be inverted is approximately the same. 
If the spatial resolution of envelope models is too low, this implies adopting
arbitrary inner boundary conditions, leading to strongly uncertain results for the values {\it and} the sign of 
$\sigma_i$. The analysis of non-radial pulsations in Cepheids based on envelope models  should thus be used with caution. Finally, note that
all results presented in this paper with the Henyey-method are obtained with complete
stellar models and a typical number of grid-points of 5000, in agreement with the required resolution estimate
based on a WKB approach.

\section{The  Riccati method}

Because of the difficulties encountered with the above-mentioned numerical method,
an independent check of the results both for high and low degree p-modes is mandatory.  
Thereby it will be of particular interest to check whether the solutions selected with the previously described matrix  method are correct and 
no physical solution has been missed. 

A method which does not suffer from resolution problems and provides
eigenvalues and eigenfunctions with prescribed accuracy 
irrespective of the order of an eigenfunction is the Riccati method.
Its application to stellar stability problems is described 
in Gautschy \& Glatzel (1990) for radial perturbations and in 
Glatzel \& Gautschy (1992) for nonradial perturbations. 
We shall only briefly summarize the basic essentials of the
approach here and refer the reader to these publications
for details.

The rapidly oscillating character in the 
central regions of the eigensolutions considered here implies
severe resolution problems due to the limited size of matrices, 
if the eigenvalue problem is solved
on the basis of a matrix eigenvalue problem. This basic difficulty,
which always occurs for high order eigenfunctions can in principle
be overcome by shooting methods, which do not rely on the inversion
of large matrices. Moreover, using a shooting method, the local
stepsize can be adjusted to match prescribed accuracy requirements.
However, for higher order boundary value problems
(the nonadiabatic nonradial stability analysis implies the solution of
a sixth order complex differential equation) shooting methods 
tend to suffer from the parasitic growth problem and are numerically
unstable. Moreover, with increasing order the number of parameters which
need to be iterated in the shooting process becomes prohibitively large.
Therefore shooting methods are usually not considered for the solution
of high order boundary value problems. 

One way to overcome the parasitic growth problem and to reduce the
parameters to be iterated to the complex eigenvalue while still
preserving the basic advantages of a shooting method consists of
transforming the linear boundary eigenvalue problem to a nonlinear
one implying a Riccati type equation. For the shooting process
we thus obtain unambiguous initial conditions at both boundaries
and the only parameter which needs to be iterated is the eigenfrequency.
In addition, integration of the nonlinear Riccati equation as 
an initial value problem is numerically stable. 
As a result, the integration of the Riccati equation provides a complex
(determinant) function of the complex frequency. Zeros of this determinant 
correspond to the eigenvalues sought.   

The latter property of the Riccati method offers an additional advantage:
eigensolutions can be determined without resorting to any initial guess.
Using matrix methods, these are usually obtained on the basis of approximations 
(e.g., the solution of the adiabatic problem) which may lead to a missing
of unexpected solutions. Using the Riccati procedure, always the full set
 of equations 
without any approximation is integrated to provide the determinant
function. By tabulating the determinant as a function of the complex
frequency, a coarse determination of the eigensolutions can be done
which subsequently may be improved by iteration. Due to the existence 
of the determinant function there is no problem of spurious eigenvalues
when using the Riccati method. 

\section{Results}

\subsection{Comparison of the results between the two methods}

We have compared results obtained with the two methods described respectively in \S 2 and \S 3 for a specific Cepheid model. Since our work is motivated by the observations of Moskalik et al. (2004), we selected a model which could describe the properties of one of the observed LMC targets where secondary periodicities were detected.
The observed Cepheid, LMC-SC2-208897, is a first overtone pulsator with a period
of $P_1$=2.42 days and absolute magnitude $M_{\rm V} \sim -3.2$. Adopting the same input physics and evolutionary code as in Alibert et al. (1999), we found that
a model with mass $m=5 \msol$, metallicity (in mass fraction) $Z=0.01$,
 $\log \, L/L_\odot$ = 3.1, $\te = 5930$ K and close to the end of central He burning
(central mass fraction of He $Y_{\rm c} \sim 4. \, 10^{-2}$) provides a good fit  for the observed magnitude and period of
LMC-SC2-208897. The calculated first overtone period is unstable with a period $P_1$=2.43 days and the absolute magnitude is $\mv$ = -3, in excellent agreement with the observed values  (see Alibert et al. 1999 for the determination of absolute magnitudes). 

\begin{figure}
\psfig{file=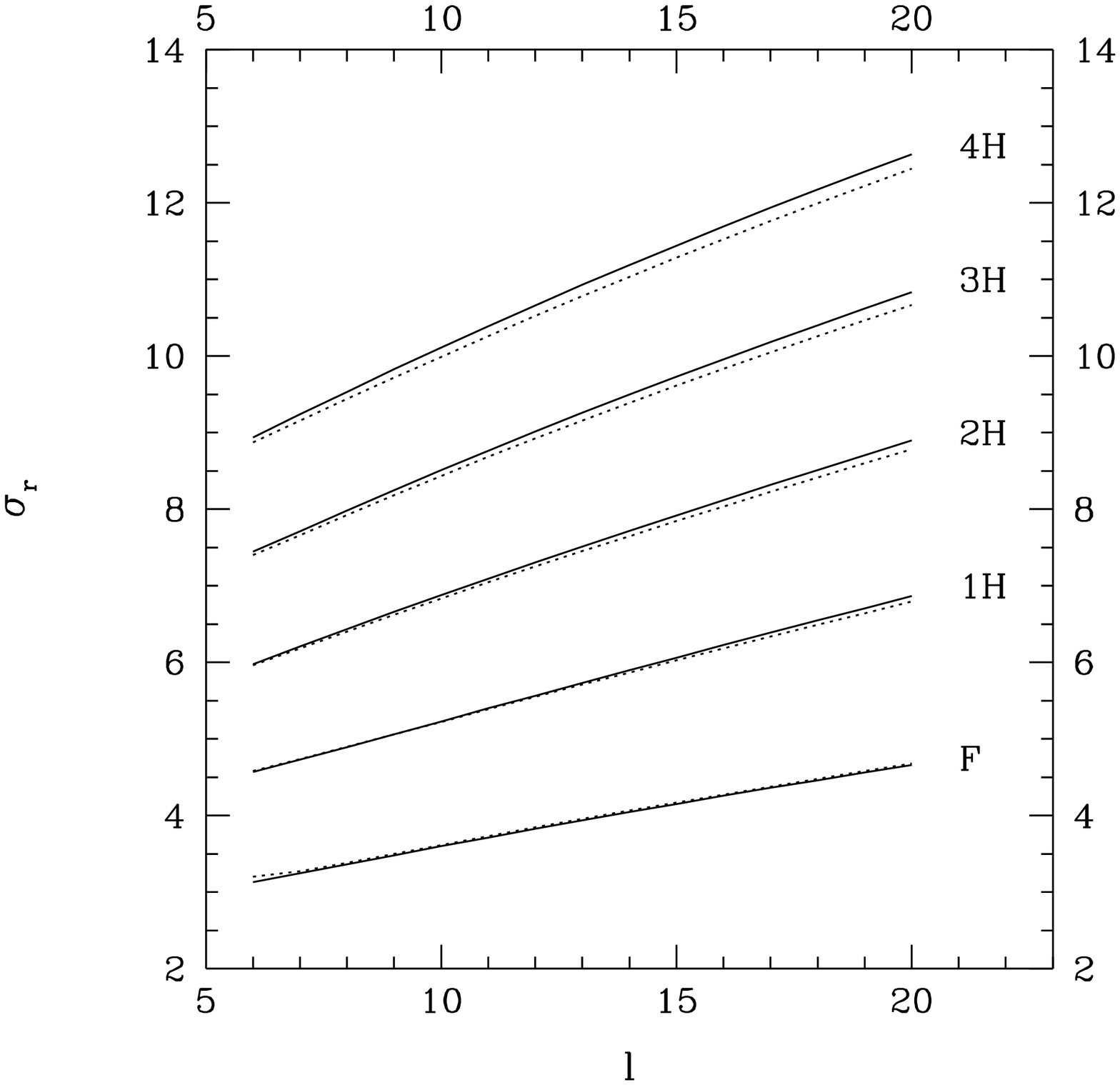,height=88mm,width=88mm}
\psfig{file=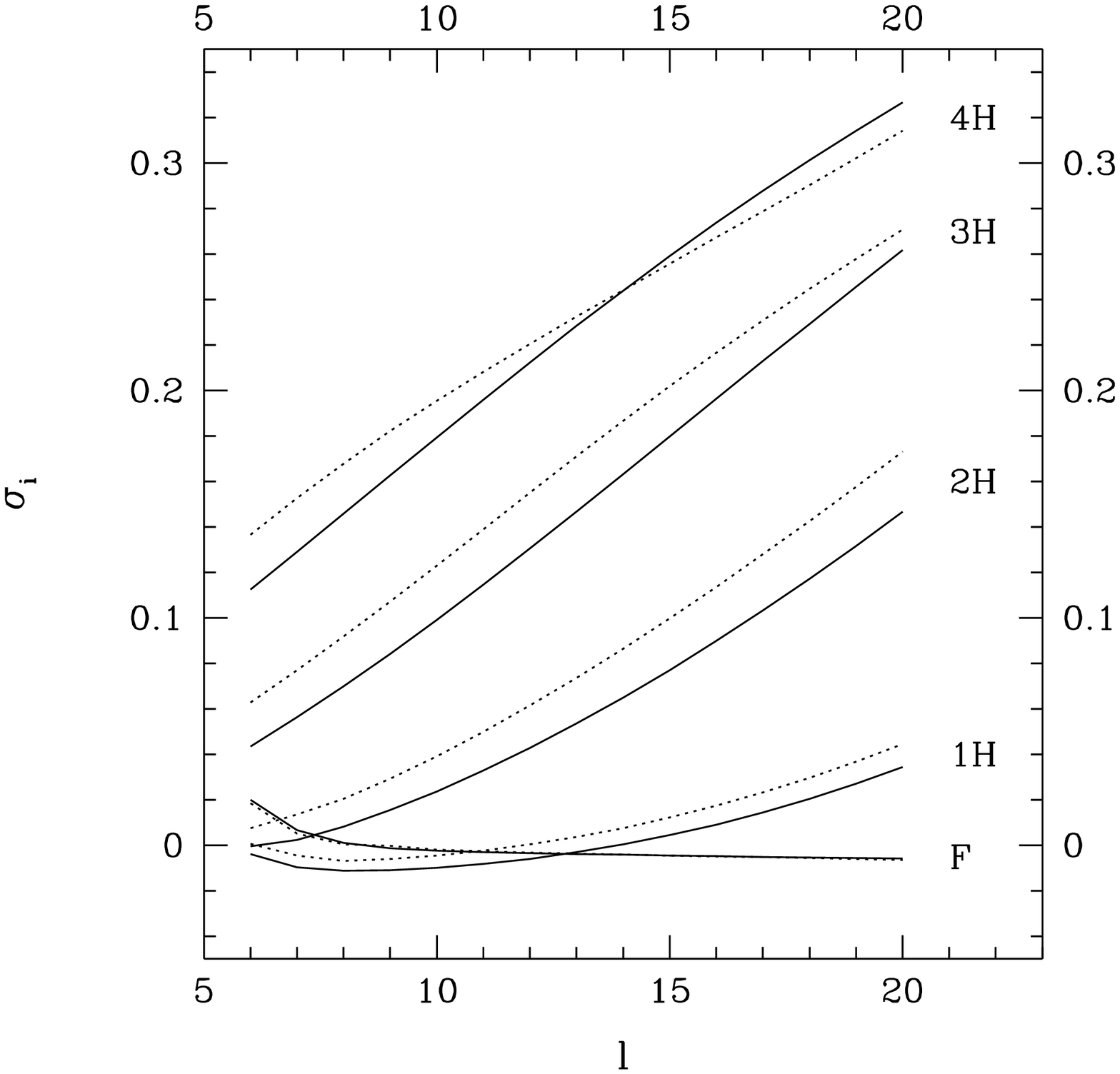,height=88mm,width=88mm}
\caption[]{Comparison, for our selected Cepheid model ($m=5M_\odot$, $\log L/L_\odot =3.1$,
 $T_{\mathrm{eff}}=5930 \, \mathrm{K}$) between the results obtained with the Henyey method (see \S 2, dotted lines) and the Riccati method (see \S 3, solid lines). The upper panel displays
the real part of eigenfrequencies of p-modes as a function of the degree $l$
and the lower panel the imaginary part, in units of  $ \sigma_0$. Each curve is labelled, on the right hand side, according to the modal classification. Note that for the fundamental mode,
the curves corresponding to the values of $\sigma_i$ obtained with both methods are indistinguishable.}
\label{fig3}
\end{figure}

Excellent agreement is found 
between the two methods for the periods and growth rates of
radial modes (see Table 1). For nonradial modes with degree $l \ge 6$, 
Fig. \ref{fig3} compares the values of $\sigma_r$ and $\sigma_i$
obtained  with both methods, up to $l$=20. The values of $\sigma_r$ agree within less than 2\%.
Differences are found for the values of $\sigma_i$ for overtones,
whereas the agreement is excellent for the fundamental mode (see Fig.  \ref{fig3}). Values of $\sigma_r$ and $\sigma_i$ for a few nonradial modes
are also given in Table 1.

\begin{table}
\caption{Linear stability analysis results for a Cepheid model with mass $m=5 \msol$, metallicity (in mass fraction) $Z=0.01$, $\log \, L/L_\odot$ = 3.1, $\te = 5930$ K.
The real part $\sigma_r$ and imaginary part $\sigma_i$ of the eigenfrequencies are normalised to $ \sigma_0  \, \sim 1.42 \, 10^5 \ {\rm cgs}$. Results obtained with the Riccati (Ric.) and the Henyey (Heny.) methods are given. Positive values of $\sigma_i$ indicate stable modes.
}		
\begin{tabular}{llccccc}
\hline\noalign{\smallskip} 
$l$ & mode & $\sigma_r$ &  $\sigma_i$ & $\sigma_r$ & $\sigma_i$ & Period \\
      &            &  (Ric.)    &  (Ric.)        & (Heny.)   &   (Heny.)  & (days) \\
\noalign{\smallskip}
\hline
0  &   F &2.99& -3.8 $10^{-4}$ &3.02&-4.4 $10^{-4}$&3.43 \\
  &  1H  &4.23&-6.3 $10^{-3}$ &4.25&-7.5 $10^{-3}$&2.43 \\
  &  2H  &5.40&-2.7 $10^{-3}$ &5.40&-4.2 $10^{-3}$&1.92 \\
 &&&&&& \\
6 & F   &3.13&2.0 $10^{-2}$ &3.20&1.8 $10^{-2}$&3.24 \\
   & 1H &4.57&-3.9 $10^{-3}$ &4.58&-5.0 $10^{-4}$&2.26 \\
   & 2H &5.98&-1.1 $10^{-3}$ &5.96&7.5 $10^{-3}$&1.74 \\
&&&&&& \\
10 & F &3.60&-2.5 $10^{-3}$ &3.61&-1.9 $10^{-3}$&2.87 \\
     & 1H &5.23&-9.8 $10^{-3}$ &5.22&-5.2 $10^{-3}$&1.98 \\
     & 2H &6.88&2.4 $10^{-2}$ &6.83&3.8 $10^{-2}$&1.52 \\
\hline
\noalign{\smallskip}

\hline \hline
  \end{tabular}
  \label{table1}
  \end{table}

For the above-mentioned effective temperature  and luminosity, a variation of the  mixing length, between 
one pressure scale height $H_{\rm P}$ and 2$\, \times \, H_{\rm P}$ or of
the metallicity between $Z$=0.01 and $Z$=0.02 have only a slight influence: $\sigma_r$
changes by less than 2.5\% and  $\sigma_i$ by less than 8 \% (if $|\sigma_i| \simgr 5. 10^{-3}$). Larger differences can appear, however, for very small values of $\sigma_i$ ($|\sigma_i| \simle 5. 10^{-3}$), when using the matrix method described in  \S 2.

\begin{figure}
\psfig{file=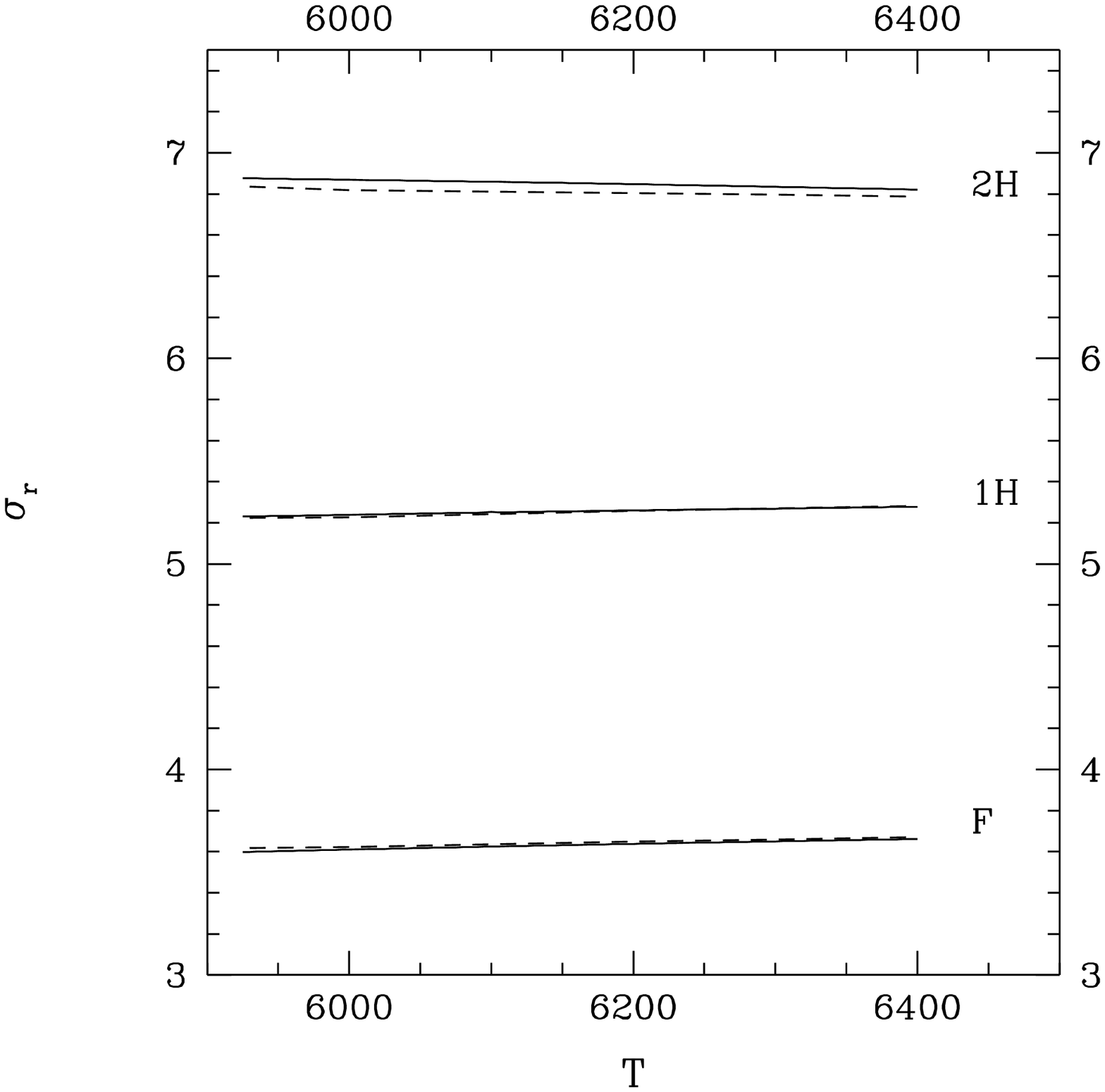,height=88mm,width=88mm}
\psfig{file=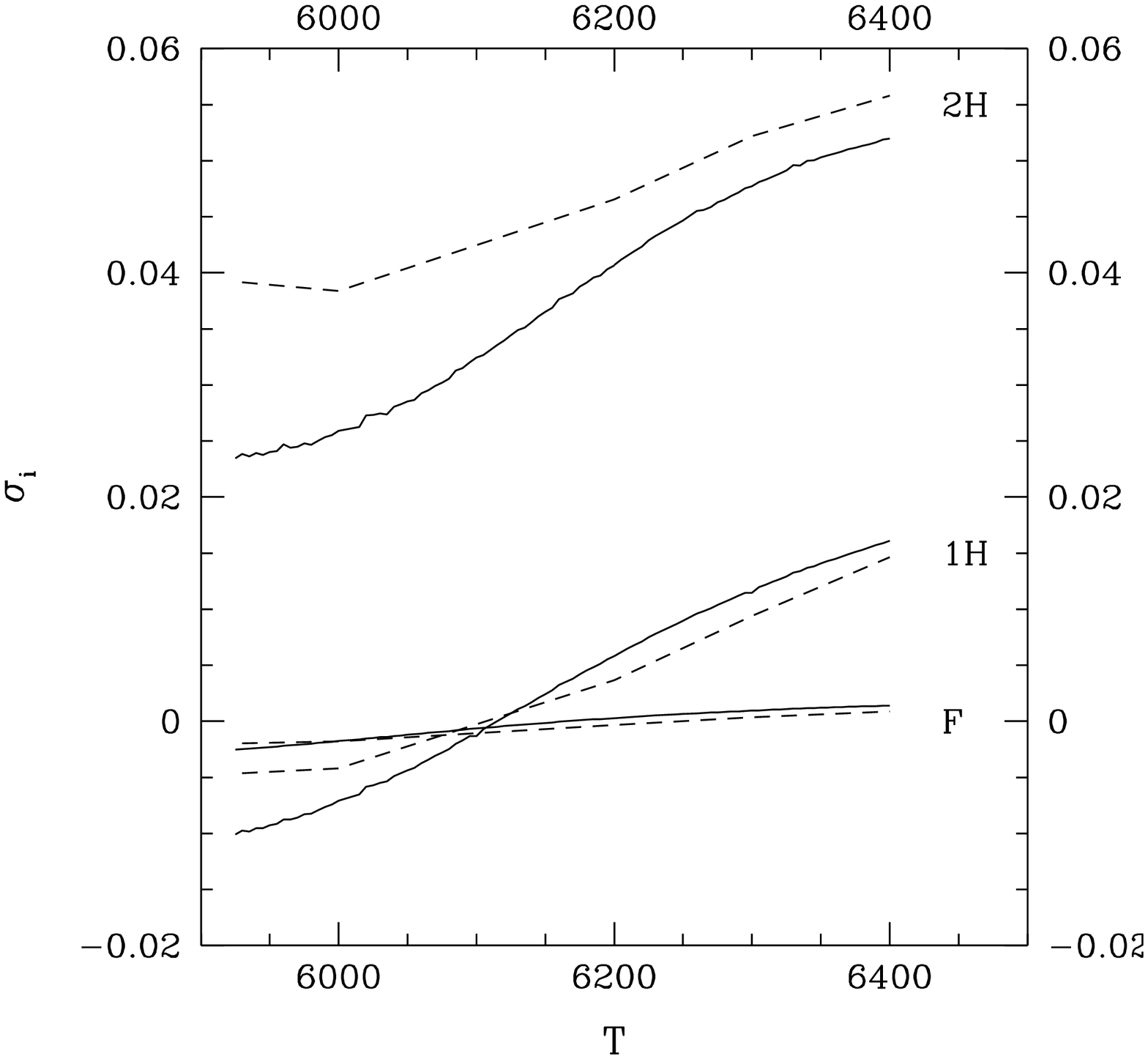,height=88mm,width=88mm}
\caption{Variation of $\sigma_r$ and $\sigma_i$, in units of $\sigma_0$, as a function of effective temperature for models with $m=5M_\odot$,  $\log L/L_\odot =3.1$ and
$Z$=0.01. Different p-modes with degree $l$= 10 are displayed, as indicated on the figure.
The solid line
corresponds to the Riccati method and the dashed line to the
Henyey method.}
\label{fig4} 
\end{figure}

We have also compared the results obtained with both methods  for different effective temperatures, adopting the same mass and luminosity as the above-mentioned Cepheid model. 
Fig. \ref{fig4} 
shows the results for the fundamental mode, 1H and 2H
with degree $l$=10. 
Here again, excellent agreement is found for the values of $\sigma_r$, but differences appear for the values
of $\sigma_i$ for the overtones. In general, both methods agree on the  sign of $\sigma_i$, providing the same results concerning the stability 
properties of modes, except for very small values of $\sigma_i$ ($|\sigma_i| \simle 5. 10^{-3}$).

Both for low and high degree modes we have tested the dependence on 
the outer boundary 
conditions of the results of the stability analysis. This is necessary,
since the outer boundary of the stellar model (the photosphere)
does not correspond to the physical boundary of the star. Therefore the
thermal and mechanical
boundary conditions are ambiguous there. In fact, both for radial and
nonradial modes we found a sensitive dependence on boundary conditions 
even within the physically admissible ones. The results shown in the 
paper, however, are based on the most conservative condition with respect
to stability. {\it I.e.}, we have chosen the condition which provides the least
unstable modes. It corresponds to the requirement of vanishing
Lagrangian pressure and temperature perturbation. All other boundary
conditions, in particular at the inner boundary, are unambiguous. 

In contrast to the thermal and mechanical outer boundary conditions
the influence of metallicity on the stability of all the models investigated 
is rather weak. Relative differences between the results for 
$Z$=0.01 and $Z$=0.02 amount to 10 per cent at maximum for $\sigma _i$.   

\subsection{Results for low-degree $p$-modes}

\begin{figure}
\psfig{file=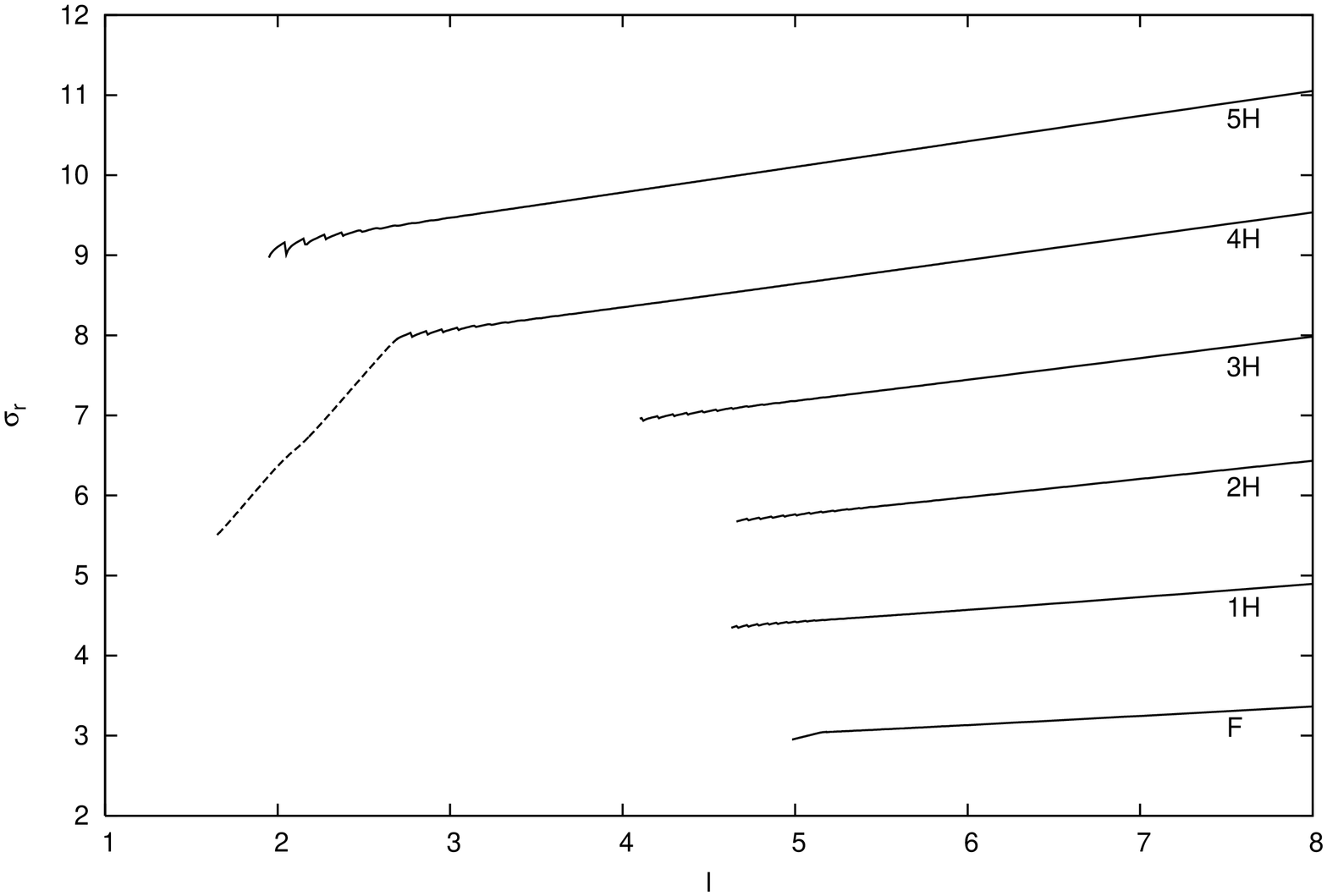,height=88mm,width=88mm,angle=-90}
\psfig{file=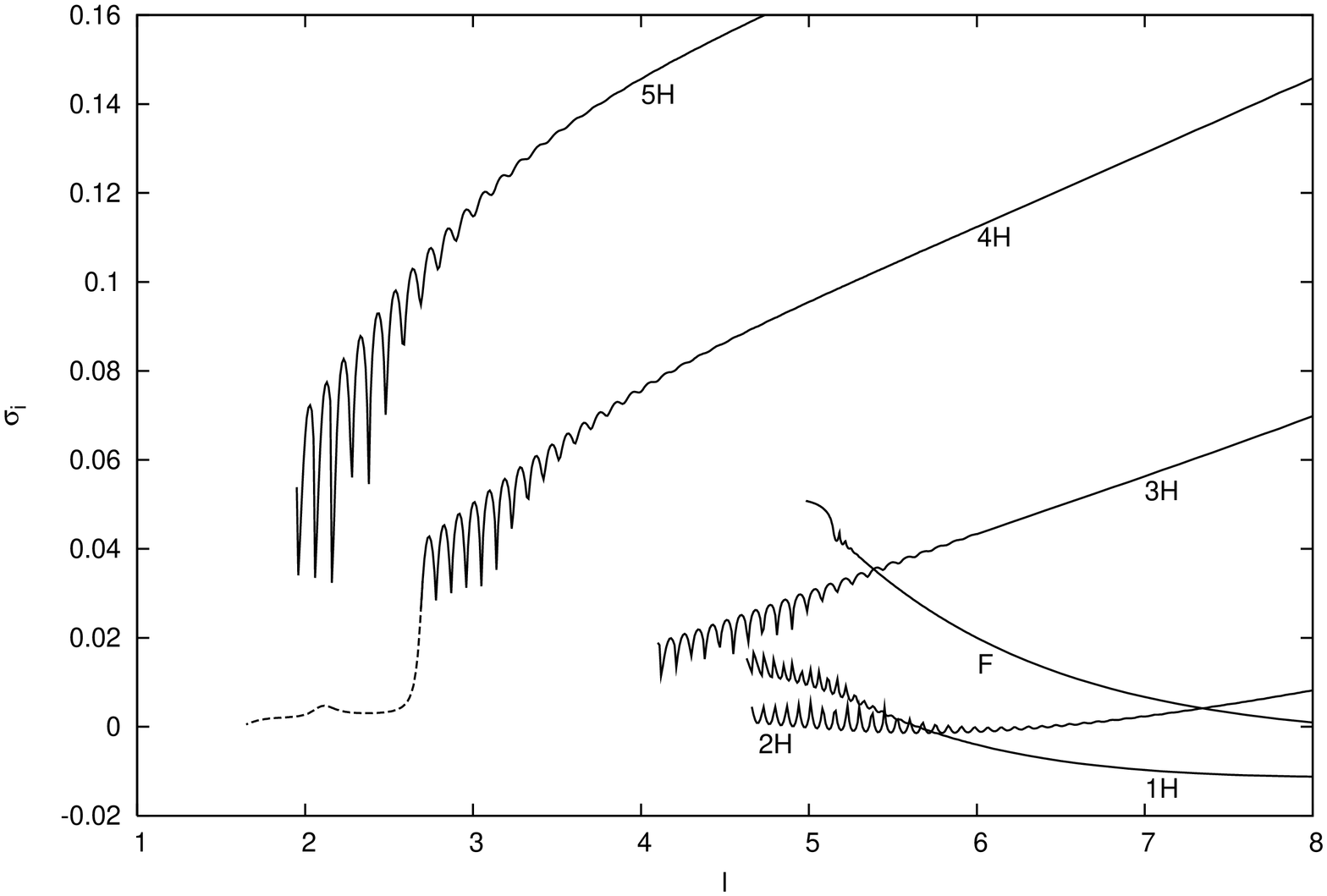,height=88mm,width=88mm,angle=-90}
\caption{Variation of $\sigma_r$ and $\sigma_i$, in units of $\sigma_0$, 
for the modes indicated as
a function of the harmonic degree for low values of $l$ and the 
Cepheid model having
$m=5M_\odot$, $\log L/L_\odot =3.1$,
 $T_{\mathrm{eff}}=5930 \, \mathrm{K}$ and $Z$=0.01.
All results are based on the Riccati method.
The bumps for low values of $l$ are due to an interaction of p- and g-modes
through avoided crossings. The dashed line indicates the run of a selected
g-mode.}
\label{fig5}
\end{figure}

Modes of degree below $l \approx 6$ are extremely difficult to find
 with matrix methods. Indeed, only high overtones
can be found ({\it e.g.} 5H for $l=2$). Some results obtained in this regime are based on the
Riccati method and shown in Fig. \ref{fig5}
. Both the frequencies of 
p- and g-modes decrease with the harmonic degree. However, while 
p-modes reach finite frequency values for $l\to 0$, the frequencies of g-modes
vanish in this limit. The stellar models considered here exhibit 
a propagation region for gravity waves close to the center of the
star which for high values of the harmonic degree is well
detached and shielded from the acoustic propagation region by 
an efficient evanescent barrier. It allows for gravity modes 
with frequencies in the range of p-modes up to quite high order. 
Since the decrease of frequencies with $l$ is stronger for the
g-modes, this leads to multiple crossings and resonances between 
p- and g-modes. Due to the efficient evanescent barrier for 
high values of $l$ the interaction of p- and g-modes is weak 
at resonances above $l \approx 6$ and leaves them unaffected.
For $l < 6$, however, the barrier becomes weaker and the resonances at the crossing of g- and
p-modes imply significant interaction and unfold into avoided crossings.
Bumps both in the real (less pronounced) and imaginary parts of the eigenfrequencies 
as a function of $l$ found in Fig. \ref{fig5} 
are parts of these 
avoided crossings. For illustration a part of the run of a g-mode
with $l$ is shown there as a dashed line and $l$ is taken as 
a real parameter. The latter allows one to follow modes 
continuously (of course, only integer values are physically meaningful).
As a consequence of the resonances,
modes with $l < 6$, in particular the so-called p-modes, should not be classified as
either p- or g-modes. Physically they rather exhibit properties of both types 
of modes.   

\section {Discussion and conclusion}

The stellar model whose parameters are thought to match the properties
of the Cepheid observed by Moskalik et al. (2004) is found unstable both
with respect to radial and nonradial perturbations. Following modes 
up to $l=500$, we find the following unstable modes:  the fundamental 
p-mode  for $l=0$ and between $9 \leq l \leq 368$, the first
overtone for  $l=0$ and between $6 \leq l \leq 13$ and the second overtone
for $l =$0, 5 and 6. No nonradial instability was found below $l=5$:
Specific studies based on the Ricatti method for $l = 1$ and 2 have not revealed 
any instability in the p-mode range. Moreover, from extrapolating the curves 
in Fig. \ref{fig5} 
we do not expect instability for $l = 3$ and 4.   
We can thus conclude that if low degree p-modes with $l \leq 2$ 
were necessary to explain the observations of Moskalik et al. (2004), they cannot be attributed to a {\it linear instability} of these
modes. Rather other effects, such as, {\it e.g.}, nonlinear mode coupling, have to 
be invoked for an explanation of their excitation.  

With respect to the dependence on effective temperature of the instability
of radial and nonradial modes we find the unstable radial modes (F, 1H, 2H) to stabilize
above 6200K, for our test case luminosity $\log \, L/L_\odot$ = 3.1. The same is true for nonradial modes with $l \leq 10$. However, fundamental p-modes
with $l > 10$ may still be unstable at temperatures above 6200 K, where no radial mode
is found to be unstable. For example,  the fundamental p-mode for $l =15$ is unstable up to
6290 K, the corresponding mode for $l =60$ is unstable up to 6640 K.     
Although observationally difficult to detect, we thus conclude that it might be worthwhile to search for nonradial pulsations
with $l \geq 5$ in Cepheids. Particularly promising appear to be observations 
of objects with $T_{\mathrm{eff}} > 6200 \, \mathrm{K}$, where no radial pulsations are expected 
and -- according to theory -- only nonradial pulsations should prevail.
Observations of nonradial pulsations in Cepheids could then provide an interesting and unique  tool  to sound their inner structure.

\begin{acknowledgements}  I.B. thanks warmly the Academy of Sciences of G\"ottingen for the Gauss professorship and for supporting this project and 
the Institute of Astronomy in G\"ottingen for hospitality during completion of this work.
The calculations were performed using the computer facilities of the Centre de Calcul Recherche et Technologie (CCRT) of CEA.
\end{acknowledgements}

\end{document}